# Write-once read-many-times memory based on a single layer of Pentacene

Shengwei Shi, Junbiao Peng, Jian Lin, and Dongge Ma

*Abstract*—We realized an organic electrical memory device with a simple structure based on single layer pentacene film embedded between Al and ITO electrodes. The optimization of the thickness and deposition rate of pentacene resulted in a reliable device with an ON/OFF current ratio as high as nearly $10^6$ which was two orders of magnitude higher than previous results, and the storage time was more than 576 hrs. The current transition process is attributed to the formation and damage of the interface dipole at different electric fields, in which the current conduction showed a transition from Ohmic conductive current to Fowler–Nordheim tunnelling current. After the transition from ON state to OFF state, the device tended to remain in OFF state even when the applied voltage was removed, which indicated that the device was very promising for write-once read-many-times (WORM) memory.

## I. INTRODUCTION

HIGH capacity and low cost write-once read-many-times (WORM) memories have widespread uses in all kinds of permanent archival storage applications where the vulnerability to breakage and the relatively high cost associated with slow and power-hungry magnetic or optical disk drives are not acceptable. Compared with rewritable memories, WORM memories require no energy to refresh stored data, and are fast and easy to access, and do not require fragile and energy consuming mechanical disk drives, which are very important for handheld applications. And compared with silicon electronics, polymer and small molecular materials have attracted considerable attentions in electronics due to their envisioned applications in low cost, flexible, large area and lightweight organic electronics [1], [2]. Several kinds of polymer and small molecular materials were also found to exhibit WORM memory effects [3-6]. Forrest et al. [3] had demonstrated a WORM memory device by integrating a Si p-i-n diode with a polymer fuse. The as-fabricated device shows a high conductivity property and can be programmed to low conductivity state after applying a bias. Yang's group [4] fabricated another WORM device from a film of polystyrene containing 2NT (2-naphthalenethiol)-capped Au nanoparticles and the device exhibits an ON/OFF current ratio of $10^3$. Recently, Y. Song et al. [5], [6] also found that a WORM device with a ON/OFF current ratio as high as $10^6$ can be realized by a single layer device based on the sandwiched structure of a conjugated copolymer (PF8Eu), containing fluorine and chelated Europium complex.

In this paper, we optimized the thickness and the deposition rate of pentacene based on the results before [7], and got a reliable memory device. We also investigated the current conductive mechanism to understand the electrical transition. The device tended to remain in OFF state even when the applied voltage was removed, showing a great potential in WORM memory applications.

## II. EXPERIMENTAL

For the studied devices, the thin film of pentacene (purchased from Aldrich and used without further purification) was fabricated by thermal deposition under a vacuum of $10^{-4}$ Pa onto the patterned indium tin oxide (ITO, with sheet resistance of 100Ω/□) cathode. The patterned ITO was precleaned and then exposed to $O_2$ plasma for 4 min. A 200 nm Al electrode was finally deposited on the top of pentacene thin film in the same vacuum system. The vertical overlap between ITO and Al electrodes as the size of the devices is $3\times 3$ mm$^2$. The optimized deposition rate was 0.1 nm/s for pentacene and 0.8~1 nm/s for Al electrode. The current-voltage (I-V) characteristics were measured by a computer-controlled sourcemeter (Keithley 2400) at room temperature under ambient conditions, and the Al was defined as anode here in all electrical measurements.

## III. RESULTS AND DISCUSSION

The deposition rate of pentacene can greatly affect the current transition, and the lower one will get current transition, while the higher one will help to stabilizing the device performance [7]. There must be an optimal one which can combine the advantages of both two, and we get such a simple and actual one layer device with the optimal deposition rate of 0.1 nm/s. Fig. 1 gives the typical I-V characteristics of ITO/pentacene (90 nm,

Manuscript received October 27, 2008. This work was supported in part by the National ''973'' Project of China (2002CB613405), and the National Natural Science Foundation of China (50573024 and 50433030), and the Key Project of Chinese Ministry of Education (104208).

Shengwei Shi was with South China University of Technology, Guangzhou 510640, P. R. China, He is now with Institut de Physique et de Chimie des Matériaux de Strasbourg, Strasbourg, 67034, France. (e-mail: hbhmfly@gmail.com).

Junbiao Peng is with South China University of Technology, Guangzhou 510640, P. R. China.

Jian Lin and Dongge Ma are with Changchun Institute of Applied Chemistry, Chinese Academy of Sciences, Changchun 130022, P. R. China.

0.1 nm/s)/Al (200 nm), in which the arrow shows the voltage sweep direction. The voltage is swept from 0 V to 15 V, and then back to 0 V. As is shown that the device can be suddenly transited from the high conductive state (ON) to the low conductive state (OFF) by applying a voltage of about 9.4 V, and when it changes to OFF state, it will be permanent and steady in the low conductive state. An ON/OFF current ratio as high as more than $10^5$ has been achieved. This feature promises a low misreading rate by precise control over the ON and OFF states.

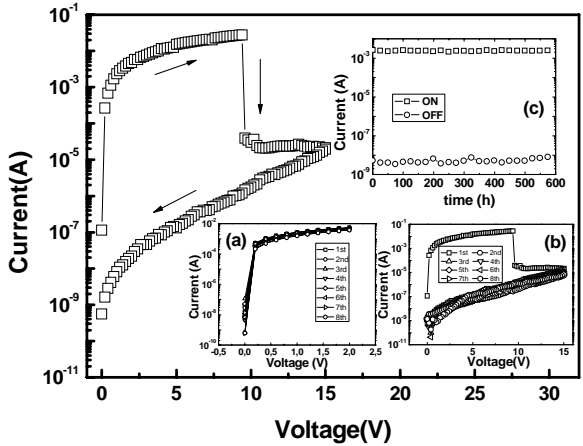

Fig. 1 Typical current-voltage (I-V) characteristic of ITO/pentacene (90 nm)/Al (200 nm) in which the arrow shows the direction of voltage sweep. Inserted (a) and (b) are the I-V characteristics for ON and OFF states of the device with 8 times continuous voltage sweep cycle; (c) currents of ON and OFF states as a function of time under a constant voltage of 1.0 V (stored in air).

Inserted (a) and (b) in Fig. 1 are the I-V characteristics for ON and OFF states of the device with 8 times continuous voltage sweep cycle. For the ON state, the sweep voltage is between 0 V and 2 V, and the current can be measured very stably at a constant low voltage as long as the current transition has not happened. For the OFF state, the sweep voltage is between 0 V and 15 V. After the current transition of the fist voltage sweep, the device will stay in the OFF state with the next voltage sweep cycle. Actually, the voltage sweep cycles can be more. Although we have no experiment results for the device during a long-term continuous operation currently; we believe that the device must have a good retention time because it shows a very stable I-V characteristic. And we have investigated on the storage time for the device stored in the air. Inserted (c) shows currents of ON and OFF states as a function of time for the device under a constant voltage of 1.0 V. The device shows a more excellent reliability with a storage time of at least 576 hours in the air than before [7]. For the measurement here, the current of the pristine device in its ON state was measured by applying 1 V bias. The measurement in the ON state was then made applying a suitable bias, which induced the OFF state, and then probed it by measuring current under a small voltage of 1 V. The current in both ON and OFF state here were measured in air every 24 hours. The results show that the device has an excellent reliability even in air ambient conditions, which is very important for practical applications.

We also investigated the effect of the thickness of pentacene on current transition, which was given in Fig. 2. When the thickness was too thin (20 nm), the devices showed electrical short, and when it was too thick (more than 150 nm), the devices showed low current injection without current transition. For the thickness range of 30-120 nm, the devices showed current transition. It is also interesting that the current level of ON state is nearly independent to the thickness of pentacene, while the current of OFF state decreases with the thickness increase. Obviously, the ON/OFF ratio is mainly determined by OFF current, but not by ON current. The inserted in Fig. 2 showed ON/OFF ratio as a function of applied voltage for the devices with different thickness of pentacene. It is easy to see that the ON/OFF ratio is greatly related with the thickness of pentacene, and the optimized thickness for pentacene is 120 nm. For the optimized device, the ON/OFF ratio can be as high as nearly $10^6$ with a lower critical voltage of 8.2 V.

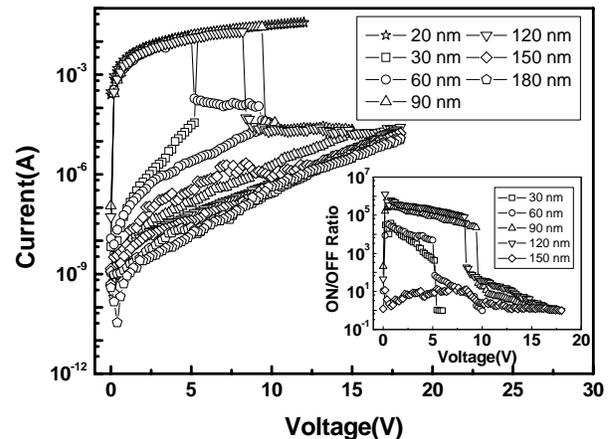

Fig. 2 I-V characteristics of the devices with different thickness of pentacene layer, inserted is ON/OFF ratio as a function of applied voltage.

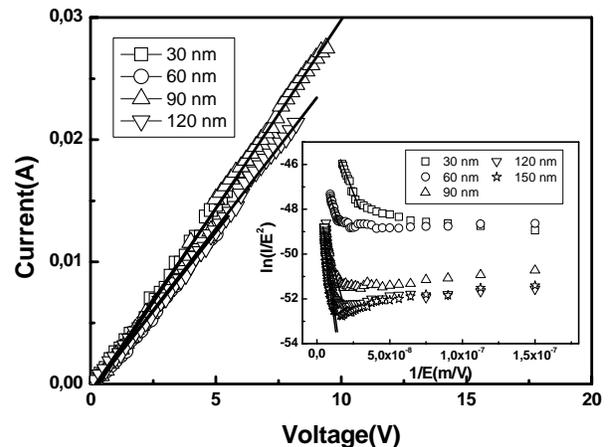

Fig. 3 I-V characteristics of ON for the devices with different thickness of pentacene layer, inserted is for OFF state.

To understand the conduction mechanism of the WORM memory device, for the devices with different thickness of pentacene layer, the I-V characteristics in both ON and OFF states are studied by using different current transport models, which were shown in Fig. 3. For the ON state, a straight linear relation in coordinates I versus V is observed for all the devices, which suggests that the current is an Ohmic conduction. For the OFF state, as shown in the inserted of Fig. 3, a linear relation of $\ln(I/E^2)$ versus $1/E$ was obtained, which suggests that the current in OFF-state is dominated by the Fowler-Nordheim tunnelling model [8]. Therefore, this observation suggests that after an electrical transition, the change of the current conduction from the ON state to OFF state is the transition from the Ohmic conduction to the injection limited current.

It is well know that when depositing Al on top of pentacene layer with a high deposition rate, the Al will easily penetrate into pentacene layer and form metallic filaments [9]. Choi et al considered that the ON state independent to the area is attributed to the filamentary conduction [10], and this could also be a reasonable explanation why the initial state shows linear Ohmic characteristic and the threshold voltage correlates with film thickness, but it can't explain the whole WORM process. We consider that the formation of a huge interface dipole layer at the electrode interface is the major origin here. Fig. 4 shows the AFM morphology and phase images for 90 nm pentacene film with the rate of 0.1 nm/s, and we can see the large grain size of pentacene. It is well established that metal atoms can migrate inside the organic layer during the deposition of the top electrode and such an interdiffusion phenomenon increases if the surface of organic film shows a larger gain size. A result of the metal diffusion will lead to the formation of a huge interface dipole layer at the electrode interface. The net effect of the dipole layer then makes the energies of the neighbouring organic molecules lower relative to the bulk, and the disorder in the dipole layer also leads to a significant broadening of the energies distribution of neighbouring organic sites, creating a reservoir of states for injected charges, thus greatly reduced the charge injection barrier.

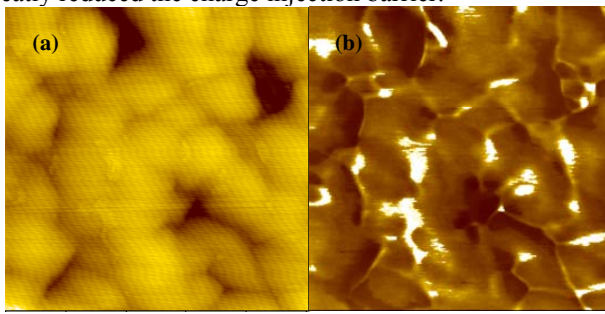

Fig.4 AFM morphology (a) and phase (b) images (1μm×1μm) for ITO/pentacene (90 nm).

Pentacene has the value of 5.0 and 3.2 eV for HOMO and LUMO, respectively [11]; the hole should be the major charge carriers for ITO/pentacene/Al. Since the interface dipole oriented in a suitable direction greatly pushes up the energy band of pentacene with respect to the vacuum level of metal, the barrier for the hole injection is significantly reduced from a Schottky contact to an Ohmic contact. Therefore, the hole injection from Al to pentacene is easier, forming the high ON state. However, the interface dipole can be destroyed due to a high space field caused by the access hole injection. As a result, the barrier height for the hole injection is then greatly increased which will lead to an abrupt decrease in current. Because of the higher injection barrier, the holes are then injected by the Fowler–Nordheim tunnelling manner, obviously the destruction of the interface dipole under a critical bias is irreversible.

For ON state, the current is Ohmic conduction, which is not affected by the thickness so much, while for OFF state, an increase in pentacene thickness indicates an increase in tunneling barriers between two electrodes which will effectively reduce OFF current [8], so ON/OFF ratio increases with the thickness. For the larger thickness, a higher critical voltage is needed to overcome the big barrier for current transition. If the thickness is too large, the device will show low current from start to end without current transition, so there is an optimal thickness for device performance, which is 120 nm for pentacene here.

## IV. CONCLUSION

In summary, we observed an electrical memory phenomenon in device of ITO/pentacene/Al, and the optimized deposition rate and thickness for pentacene are 0.1 nm/s and 120 nm, respectively. The device exhibited two conductive states with an ON/OFF ratio as high as nearly $10^6$, and showed an excellent reliability with a storage time more than 576 hours. The reduction of the barrier height caused by a large interface dipole and the damage of the interface dipole under a critical bias voltage has been used to explain the transition processes. The current conduction showed a transition from Ohmic conductive current to Fowler–Nordheim tunnelling current during the device switching. Our results indicate that the device is a very potential memory in WORM application, which will open up a new application field for the pentacene organic semiconductor.

### REFERENCES

[1] S. R. Forrest, "The path to ubiquitous and low-cost organic electronic appliances on plastic," Nature, vol. 428, pp. 911-918, 2004.
[2] J. M. Shaw and P. F. Seidler, "Organic electronics: Introduction," IBM J. Res. Dev., Vol. 45, pp. 3-9, 2001.
[3] S. Moller, C. Perlov, W. Jackson, C. Taussig, and S. R. Forrest, "A polymer/semiconductor write-once read-many-times memory," Nature, vol. 426, pp. 166-169, 2003.
[4] J. Ouyang, C. W. Chu, D. Sieves, and Y. Yang, "Electric-field-induced charge transfer between gold nanoparticle and capping 2-naphthalenethiol and organic memory cells," Appl. Phys. Lett., vol. 86, pp. 123507, 2005.
[5] Y. Song, Q. D. Ling, C. Zhu, E. T. Kang, D. S. H. Chan, Y. H. Wang, and D. L. Kwong, "Memory Performance of a Thin-Film Device Based on a Conjugated Copolymer Containing Fluorene and Chelated Europium Complex," IEEE Electron Device Lett., vol. 27, no. 3, pp. 154-156, 2006.
[6] Y. Song, Y. P. Tan, E. Y. H. Teo, C. X. Zhu, D. S. H. Chan, Q. D. Ling, K. G. Neoh, and E. T. Kang, "Synthesis and memory properties of a conjugated copolymer of fluorene and benzoate with chelated europium complex," J. Appl. Phys., vol. 100, pp. 084508, 2006.
[7] J. Lin and D. Ma, "The morphology control of pentacene for write-once-read-many-times memory devices," J. Appl. Phys., vol. 103, pp. 024507, 2008.
[8] S. M. Sze, Physics of Semiconductor Devices, 2nd ed. New York: Wiley, 1981.
[9] D. Tondelier, K. Lmimouni, D. Vuillaume, C. Fery, and G. Haas, "Metal/organic/metal bistable memory devices," Appl. Phys. Lett., vol. 85, pp. 5763-5765, 2004.
[10] S. Choi, S. H. Hong, S. H. Cho, S. Park, S. M. Park, O. Kim, and M. Ree, "High-Performance Programmable Memory Devices Based on Hyperbranched Copper Phthalocyanine Polymer Thin Films," Adv. Mater., vol. 20, pp. 1766-1771, 2007.
[11] S. Shi and D. Ma, "A pentacene-doped hole injection layer for organic light-emitting diodes," Semicond. Sci. Technol. Vol. 20, pp. 1213-1216, 2005.